\documentclass[%
 reprint,
 amsmath,amssymb,
 aps,
]{revtex4-2}

\usepackage{graphicx}
\usepackage{dcolumn}
\usepackage{bm}

\usepackage{braket}

\begin{document}

\preprint{APS/123-QED}

\title{Microwave electrometry with Rydberg atoms in a vapor cell using microwave amplitude modulation}
\author{JIANHAI HAO}
\affiliation{ 
 School of Physical Sciences, University of Chinese Academy of Sciences, Beijing 100049, China
}%

\author{FENGDONG JIA}%
\email{fdjia@ucas.ac.cn}
 \altaffiliation[Author to whom correspondence should be addressed: ]{{fdjia@ucas.ac.cn}}
 
\affiliation{ 
 School of Physical Sciences, University of Chinese Academy of Sciences, Beijing 100049, China
}%

\author{YUE CUI}
\affiliation{
School of Physical Sciences, University of Chinese Academy of Sciences, Beijing 100049, China
}%

\author{YUHAN WANG}
\affiliation{
School of Physical Sciences, University of Chinese Academy of Sciences, Beijing 100049, China
}%

\author{FEI ZHOU}
\affiliation{
School of Physical Sciences, University of Chinese Academy of Sciences, Beijing 100049, China
}

\author{XIUBIN LIU}
\affiliation{
School of Physical Sciences, University of Chinese Academy of Sciences, Beijing 100049, China
}

\author{JIAN ZHANG}
\affiliation{
Institute of Nuclear and New Energy Technology, Collaborative Innovation Center of Advanced Nuclear Energy Technology, Key Laboratory of Advanced Reactor Engineering and Safety of Ministry of Education, Tsinghua University, Beijing 100084, China
}

\author{FENG XIE}
\affiliation{
Institute of Nuclear and New Energy Technology, Collaborative Innovation Center of Advanced Nuclear Energy Technology, Key Laboratory of Advanced Reactor Engineering and Safety of Ministry of Education, Tsinghua University, Beijing 100084, China
}

\author{ZHIPING ZHONG}
\affiliation{
School of Physical Sciences, University of Chinese Academy of Sciences, Beijing 100049, China
}%
\affiliation{
CAS Center for Excellence in Topological Quantum Computation, University of Chinese Academy of Sciences, Beijing 100190, China
}%


\begin{abstract}
We have theoretically and experimentally studied the dispersive signal of the Rydberg atomic electromagnetically induced transparency (EIT) - Autler-Townes (AT) splitting spectra obtained using amplitude modulation of the microwave (MW) field. In addition to the two zero-crossing points, the dispersion signal has two positive maxima with an interval defined as the shoulder interval of the dispersion signal $\Delta f_{\text{sho}}$. The relationship of MW field strength $E_{\text{MW}}$ and $\Delta f_{\text{sho}}$ are studied at the MW frequencies of 31.6 GHz, 22.1 GHz, and 9.2 GHz respectively. The results show that $\Delta f_{\text{sho}}$ can be used to character the much weaker $E_{\text{MW}}$ than the interval of two zero-crossing points $\Delta f_{\text{zeros}}$ and the traditional EIT-AT splitting interval $\Delta f_{\text{m}}$, the minimum $E_{\text{MW}}$ measured by $\Delta f_{\text{sho}}$ is about 30 times smaller than that by $\Delta f_{\text{m}}$. As an example, the minimum $E_{\text{MW}}$ at 9.2 GHz that can be characterized by $\Delta f_{\text{sho}}$ is 0.056 mV/cm, which is the minimum value characterized by frequency interval using vapour cell without adding any auxiliary fields. The proposed method can improve the weak limit and sensitivity of $E_{\text{MW}}$ measured by spectral frequency interval, which is important in the direct measurement of weak $E_{\text{MW}}$.
\end{abstract}

\maketitle

%

\section{INTRODUCTION }

Quantum sensing generally refers to the use of quantum systems, quantum properties, or quantum phenomena to measure physical quantities \cite{1,2,3,4,5,6,7,8,9,10,11,12,13,14,15,16,Quantum_sensing_2017,Wang_23,Cui2023pra}. Rydberg atoms are highly excited atoms, which have a long lifetime and large polarizability, and the energy level interval between adjacent Rydberg states is in the microwave (MW) band. This makes Rydberg atoms sensitive to MW and makes them ideal quantum sensors for precise MW measurements. Based on the Rydberg electromagnetically induced transparency (EIT), the addition of MW can cause the splitting of the Rydberg atomic energy level, which in turn causes the Autler-Townes (AT) splitting of the EIT. Using the EIT-AT effect, MW strength can be converted into optical frequency outputs \cite{4}. It is well known that optical frequency can be absolutely and directly measured. Therefore, the measurement of microwave electric field based on optical frequency interval is the most accurate. At present, the minimum resolution based on the traditional EIT-AT method in the room temperature atomic vapor cell is $\sim$5 mV/cm \cite{4}. The measurement of electric field intensity based on EIT-AT splitting is traceable and widely used to calibrate weak MW electric fields.  Therefore, how to expand the method of measuring weak fields based on EIT-AT splitting is of great significance.

 Various methods have been developed to extend the lower limit of EIT-AT splitting measurement of electric fields, including cold atom method \cite{liao_2020,Zhou_2023}, detuning method \cite{simon_2016}, auxiliary MW field method \cite{Jia_2021,Yuan_2022}, amplitude modulation (AM) dispersion method \cite{liu_2021}, and three-photon excitation method \cite{Ripka_2022}. Given the nature of cold atoms with their low temperature and lower dephasing rate, the line width of the Rydberg EIT can be reduced significantly, making it more conducive to precision measurements. Kai-Yu Liao \emph{et al}. narrowed the Rydberg electromagnetic induction absorption (EIA) linewidth in a cold atom sample to 500 kHz, and further used  
EIA-AT splitting to achieve a measurable minimum MW field intensity resolution of 100 $\mu$V/cm in the linear region of EIA-AT splitting \cite{liao_2020}. Using EIT-AT splitting in cold atomic samples, Fei Zhou \emph{et al}. achieved a lower limit of 222 $\mu$V/cm for the linear region of EIT-AT splitting \cite{Zhou_2023,zhou_2022}. Compared to thermal atomic vapor cells, laser-cooled atoms require expensive and complex vacuum equipments and laser cooling systems, and the long preparation cycle of cold atoms is not conducive to the development and maintenance of subsequent products. Therefore, the Rydberg MW electric field sensor based on the vapor cell is still a hot research topic in this field. Simons \emph{et al}. changed the microwave frequency from resonance to detuning, and the weak MW field could be inferred from the generated non-resonant EIT-AT splitting interval \cite{simon_2016}. Compared with the measurement of MW field by EIT-AT splitting during resonance, this method can improve the measurement sensitivity by about 2 times \cite{simon_2016}. Feng-Dong Jia \emph{et al}. made the Rydberg EIT-AT splitting generated by the signal MW field more significant by adding an auxiliary MW field with a frequency different from that of the signal MW field, which in turn extended the lower limit of the direct SI traceable electric field strength to two orders of magnitude lower \cite{Jia_2021}. Jin-peng Yuan \emph{et al}. improved the sensitivity of the measurement by about two orders of magnitude by using an auxiliary field at the same frequency as the signal MW field \cite{Yuan_2022}. Xiu-bin Liu \emph{et al}. used the amplitude modulation of MW to resolve EIT-AT splitting in weak MW fields, and measured the intensity of the MW field using spectral frequency interval between two zero-crossing points near the EIT resonance in the dispersion signal, increasing the minimum distinguishable MW strength by a factor of about 6 \cite{liu_2021}. Fabian Ripka \emph{et al}. obtained an EIT linewidth of 500 kHz in a cesium atomic vapor cell using collinear three-photon excitation to eliminate residual Doppler effects, advancing the lower limit of the EIT-AT split linear region in an atomic vapor cell toward the weak field region by one order of magnitude \cite{Ripka_2022}. Nikunjkumar Prajapati \emph{et al}. compared in detail the advantages and disadvantages of two-photon excitation and three-photon excitation and found that these narrow line features do not correspond directly to the best sensitivity \cite{Nikunjkumar_2022}. Xiubin Liu \emph{et al}. proposed that the method of MW electric field AM has the advantages of strong operability and adaptability, and can be used for two-photon and three-photon excitation to further expand the measuring range of the EIT-AT splitting region \cite{liu_2021}.

This paper can be regarded as a further study of the work of Xiubin Liu \emph{et al}. on the use of AM of the MW field to obtain dispersive signals and then measure MW weak fields \cite{liu_2021}. We investigate in detail the spectral characteristics of the dispersive signal obtained by the amplitude modulation of the MW field theoretically and experimentally, respectively. The dispersion signal obtained by modulating the MW amplitude not only has two zero-crossing points but also has two positive extreme values. The effect of modulation depth on the profile of the dispersion signal is discussed, and the interval based on two maxima (shoulder interval of the dispersion signal) is discussed to characterize the MW strength. The paper is organized as follows, Sec. II is a theoretical and detailed study of the effect of MW modulation depth on the dispersion spectrum using numerical calculations, and investigates the relationship between different MW strengths and the shoulder interval of the dispersive signal at the optimal MW modulation depth, and then shows the lower limits of the measurements by different methods. Sec. III will introduce the experimental system and the experimental method. Sec. IV will focus on our experimental results obtained at MW frequencies of 31.6 GHz, 22.1 GHz, and 9.2 GHz, respectively. Sec. V is the conclusion of this paper.

\section{THEORETICAL MODEL }
\begin{figure}
\includegraphics[scale={0.6}]{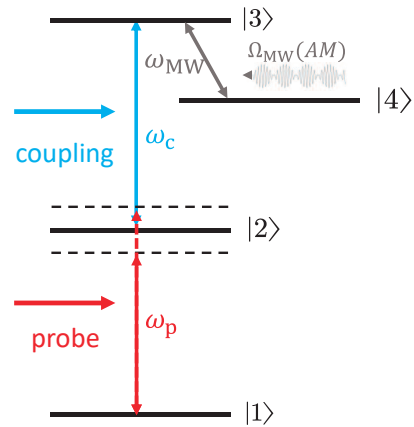}
\caption{\label{fig:1} Energy four-level diagram for MW amplitude modulation. The diagram involves four energy levels, two laser fields, and one amplitude-modulated MW field. Among them, the red arrow indicates the probe laser, $\omega_{\text{p}}$ is the angular frequency of the probe laser and it scans near the transition of $\ket{1}$ to $\ket{2}$; the blue arrow indicates the coupling laser, $\omega_{\text{c}}$ is the angular frequency of the coupling laser and it is always constant; the gray arrow tables the MW field, $\Omega_{\text{MW}}(AM)$ is the amplitude-modulated MW Rabi frequency.}
\end{figure}

The four levels model of \,$^{87}$Rb atoms driven by an amplitude-modulated microwave is shown in Figure \ref{fig:1}. The probe laser couples the ground state $\ket{1}$ to the intermediate state $\ket{2}$, and the coupling laser further couples the intermediate state $\ket{2}$ to the Rydberg state $\ket{3}$. The ground state $\ket{1}$, intermediate state $\ket{2}$ and Rydberg state $\ket{3}$ form a typical ladder type EIT configuration. To obtain the Rydberg EIT spectrum, we make the frequency $\Omega_{\text{c}}$ of the coupling laser constant and the frequency $\Omega_{\text{p}}$ of the probe laser continually scanning between the upper and lower dashed lines of the intermediate state$\ket{2}$. A MW field couples the two adjacent Rydberg energy levels $\ket{3}$ and $\ket{4}$. When the resonant MW field is strong enough, the EIT peak splits symmetrically into two, the so-called EIT-AT splitting. 

By measuring the interval $\Delta{f_\text{m}}$ of EIT-AT splitting, the MW electric field strenght $E_{\text{MW}}$, which is proportional to $\Delta{f_\text{m}}$, can be obtained \cite{Holloway_2014}:
\begin{eqnarray}\label{eq1}
E_{\text{MW}}=2\pi \frac{\hbar}{\mu} \frac{\lambda_\text{p}}{\lambda_\text{c}}\Delta{f_\text{m}},
\end{eqnarray}
where $\mu$ is the MW transition dipole moment of corresponding energy level, $\hbar$ is Planck's constant, $\lambda_p$ and $\lambda_c$ are the wavelengths of the probe laser and the coupling laser, respectively. $D$ is the Doppler mismatch factor, which explains the Doppler mismatch of the reverse transmitted the probe and coupling lasers. $D=1$ means scanning the coupling laser without considering Doppler mismatch \cite{2}. In this paper, $D=\frac{\lambda_\text{p}}{\lambda_\text{c}}$ represents scanning probe laser and the Doppler mismatch between the probe and coupling lasers is considered \cite{2}. When $\Delta f_{\text{m}}$ is smaller than the EIT linewidth $\Gamma _{\text{EIT}}$, $\Delta f_{\text{m}}$ cannot be obtained directly \cite{Holloway_2017}, i.e., the weak MW cannot be detected. In order to further directly measure weaker MW, the amplitude modulation of MW $\Omega_{\text{MW}}\text{(AM)}$ is further studied in this work.

\begin{figure*}
\includegraphics[scale=0.5]{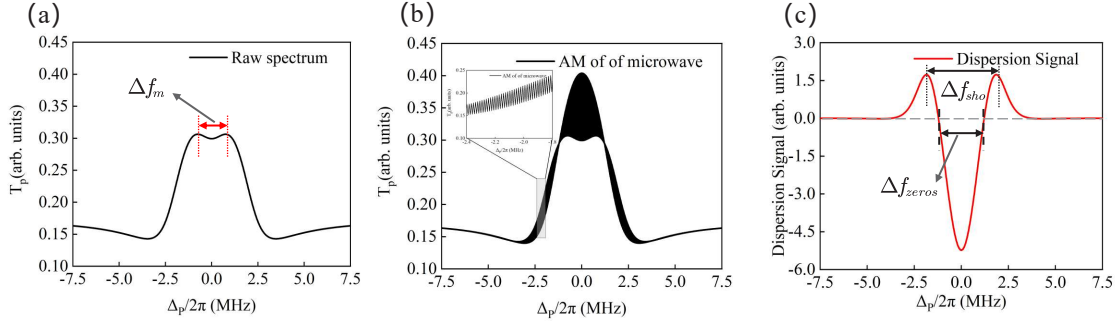}
\caption{\label{fig:2} Calculations of the spectrum of each stage in the AM of MW process. (a) Original EIT-AT splitting spectrum; (b) EIT-AT splitting spectrum after modulating MW amplitude, and modulation depth is 100\%; (c) Dispersion signal after demodulation and low-pass filtering. The calculation parameters are $\Omega _{\text{p}}=2\pi \times  1.0\,\text{MHz} $, $\Omega _{\text{c}}=2\pi \times 4.0\,\text{MHz} $ and $\Omega _{\text{MW}}=2\pi \times  2.5\,\text{MHz} $.}
\end{figure*}

Amplitude-modulated MW is performed as shown in Figure~\ref{fig:1}, which will cause the Rydberg energy levels of $\ket{3}$ and $\ket{4}$ to be periodically shifted at the modulation frequency. 
It is equivalent to the frequency modulation of the EIT-AT splitting spectrum, and further dispersion signals can be obtained. We have modeled and calculated the four-level EIT-AT splitting spectra through the ATOMIC DENSITY MATRIX package \cite{matrix_1,matrix_2}. The weak probe condition is adopted in the modeling process, and the Doppler convolution is further obtained by numerical integration \cite{Holloway_2017,zhou_2022}. The modulated MW Rabi frequency $\Omega_{\text{MW}}(AM)$ can be expressed as:
\begin{equation}\label{eq2}
\Omega _{\text{MW}}\left( AM \right) =\Omega _{\text{MW}}\cdot \frac{1}{2}\left[ \left( 2-MD \right) +MD\cdot \sin \left( \omega_{\text{am}}\cdot \Delta _{\text{p}} \right) \right], 
\end{equation}
where $\Omega_{\text{MW}}$ is the Rabi frequency of the unmodulated MW field, MD is the modulation depth (i.e., the amplitude ratio before and after modulation), $\omega_{\text{am}}$ is the amplitude modulation frequency, and $\Delta_{\text{p}}$ is the probe laser detuning.

Under the rotating wave approximation, the Hamiltonian quantities of the four energy level system can be expressed as:
\begin{equation}\label{eq3}
H=\left[ \begin{matrix}
	0&		\Omega _{\text{p}}&		0&		0\\
	\Omega _{\text{p}}&		-2\Delta _{\text{p}}&		\Omega _{\text{c}}&		0\\
	0&		\Omega _{\text{c}}&		-2\left( \Delta _{\text{p}}+\Delta _{\text{c}} \right)&		\Omega _{\text{MW}}\left( AM \right)\\
	0&		0&		\Omega _{\text{MW}}\left( AM \right)&		-2\left( \Delta _{\text{p}}+\Delta _{\text{c}}-\Delta _{{\text{MW}}} \right)\\
\end{matrix} \right], 
\end{equation}
where $ \Omega _{\text{i}}$ is the Rabi frequency, $ \Delta _{\text{i}}=\omega _{\text{i}}-\omega _{{\text{io}}}$ is the detuning of the optical field with the upper and lower energy levels of the atomic leap, $ \omega _{\text{i}}$ is the angular frequency of the optical field, and $ \omega _{{\text{i0}}}$ is the angular frequency of the resonance corresponding to the atomic leap. The subscripts ${\text{i}}$ are ${\text{p}}$, ${\text{c}}$, or ${\text{MW}}$, where ${\text{p}}$ represents the probe laser, ${\text{c}}$ represents the coupling laser, and ${\text{MW}}$ represents the target MW field.

The Hamiltonian H of the above four-level atomic system is substituted into the main equation:
\begin{equation}\label{eq4}
\dot{\rho}=\frac{\partial \rho}{\partial t}=-\frac{i}{\hbar }\left[ H,\rho \right] +\mathcal{L},
\end{equation}
where $\mathcal{L}$ is the Lindblad operator that represents the relaxation term related to the atomic decay process. Under the steady-state condition of $\dot{\rho}$, the analytical solution of the density matrix element $\rho_{12}$ can be obtained from the Eq. (\ref{eq4}). The $\rho _{12_D}$ can be further obtained after considering Doppler averaging \cite{Holloway_2017}.

Then, the magnetization $\chi$ of the medium in the cell is obtained from $\chi = \left( \frac{2\mathcal{N}_0\varepsilon_{12}}{E_\text{p}\epsilon _\text{0}} \right) \rho_{12_\text{D}}$. Let $\alpha$ be $\frac{2\pi\text{Im}\left[ \chi \right]}{\lambda_\text{p}}$. Here $\mathcal{N}_0$ is the $^{87}Rb$ atomic density in the cell, $\varepsilon _{12}$ is the transition dipole moment from the ground state $\ket{1}$ to the intermediate state $\ket{2}$, $E_\text{p}$ is the electric field amplitude of the probe laser, $\epsilon_0$ is the vacuum permittivity, $\lambda_\text{p}$ is the wavelength of the probe laser, and $\alpha$ is the Beer absorption coefficient of the probe laser transmitted in the medium. According to Beer-Lambert law, the transmission ratio of the probe laser before and after passing through the cell filled with $^{87}Rb$ atomic vapor is obtained: $T=\frac{I}{I_0}=exp\left( -\alpha l \right) $. Where $I_\text{0}$ and $I$ represent the optical signals before and after the probe laser passes through the cell respectively, and $l$ represents the length of the cell containing $^{87}Rb$ atoms. Further, we amplitude modulate the MW Rabi frequency by Eq. (\ref{eq2}) and analyze the characteristics of the EIT-AT dispersion signal. After modulating and demodulating the transmission signal $I\left( v,\Omega _{\text{MW}} \right)$ of the detection light and performing low-pass filtering, the dispersion signal associated with the probe laser spectrum is obtained.

We first discuss the effect of the modulation-demodulation process in the theoretical calculation on the EIT-AT spectrum, as shown in Figure \ref{fig:2}. Figure \ref{fig:2}(a) shows the original EIT-AT spectral signal when $\Omega _\text{p}=2\pi \times 1.0\,\text{MHz} $, $\Omega _{\text{c}}=2\pi \times  4.0\,\text{MHz} $ and $\Omega _{\text{MW}}=2\pi \times  2.5\,\text{MHz} $; Figure \ref{fig:2}(b) shows the EIT-AT signal modulated by $\Omega_{\text{MW}}$, where MD=100\%; Figure \ref{fig:2}(c) shows the dispersion signal after demodulation and low-pass filtering of the modulated signal. The EIT-AT splitting interval $\Delta f_{\text{m}}$ is 1.51 MHz as shown in Figure \ref{fig:2}(a), the interval $\Delta f_{\text{zeros}}$ between the two zero-crossing points of the dispersion signal is 2.42 MHz as shown in Figure \ref{fig:2}(c), and the shoulder interval of the dispersion signal $\Delta f_{\text{sho}}$  is 3.71 MHz as shown in Figure \ref{fig:2}(c). The spectral peak resolution of the dispersion signal is significantly higher than that of the original EIT- AT spectral signal. This indicates that the lower resolution limit of $E_{\text{MW}}$ can be further extended by measuring $\Delta f_{\text{sho}}$. 

\begin{figure}
\includegraphics[scale=0.35]{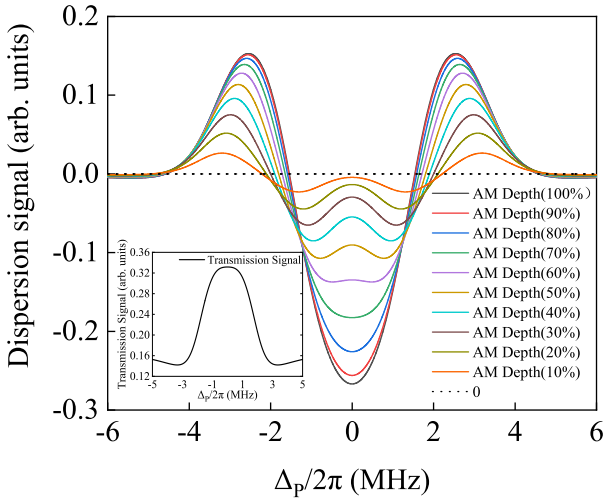}
\caption{\label{fig:5} Calculations of the dispersion signals with different modulation depths. The inset shows the original spectrum with $\Omega _{\text{MW}}=2\pi \times  2.0\,\text{MHz} $, note that the EIT-AT can not be distinguished. The calculation parameters are $\Omega _{\text{p}}=2\pi \times  1.0\,\text{MHz} $, $\Omega _{\text{c}}=2\pi \times  4.0\,\text{MHz} $ and $\Omega _{\text{MW}}=2\pi \times  2.0\,\text{MHz} $.}
\end{figure}

\begin{figure}
\includegraphics[scale=0.3]{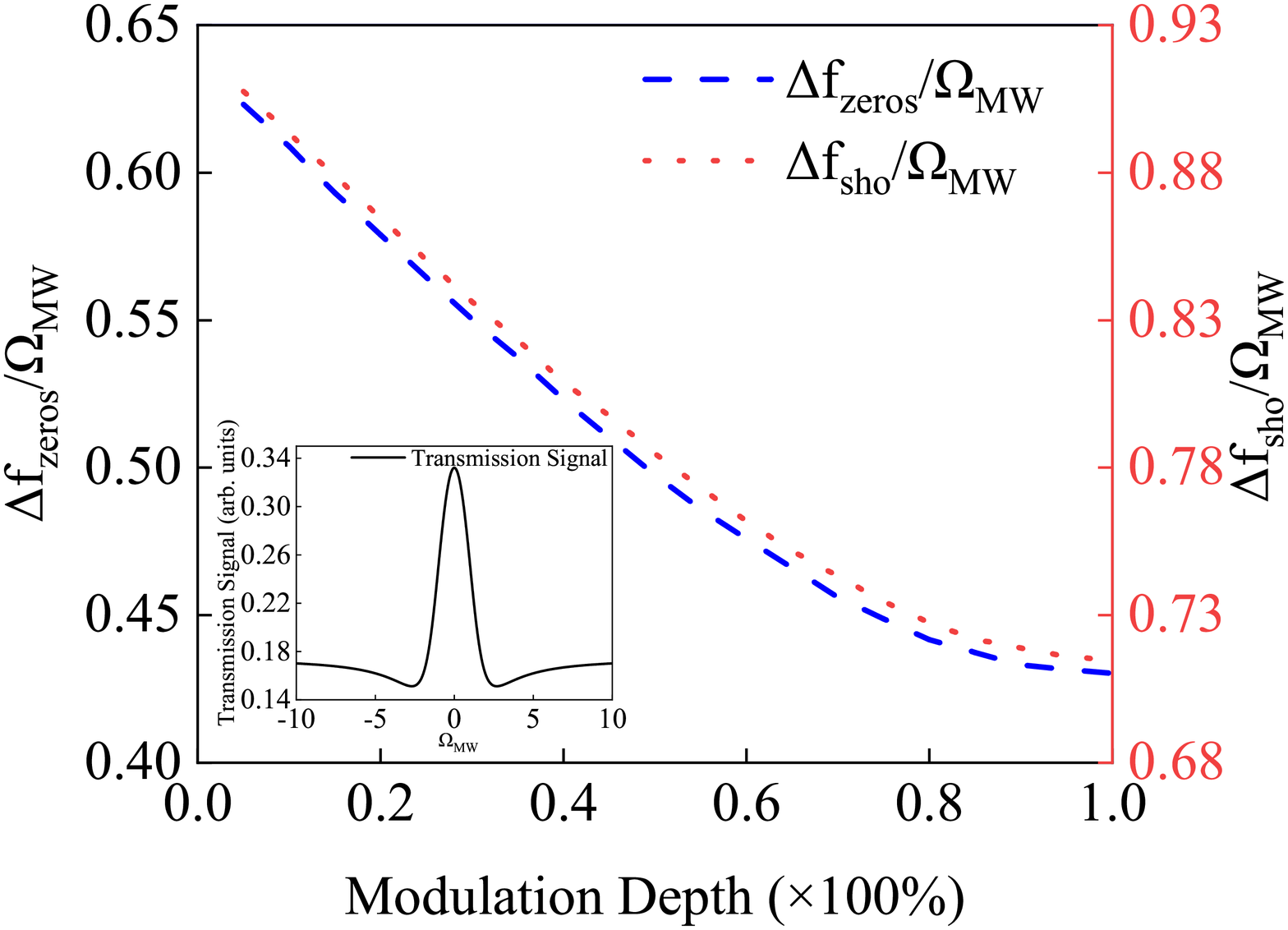}
c\caption{\label{fig:6} Calculation results of the relationship between the interval of zero crossing of dispersion signals $\Delta f_{\text{zeros}}$, the dispersion signal shoulder interval $\Delta f_{\text{sho}}$ and different modulation depths. The blue dashed line represents $\Delta f_{\text{zeros}}/\Omega_{\text{MW}}$; the red dotted line represents $\Delta f_{\text{sho}}/\Omega_{\text{MW}}$.The calculation parameters are $\Omega _{\text{p}}=2\pi \times  1.0\,\text{MHz}$, $\Omega _{\text{c}}=2\pi \times  4.0\,\text{MHz} $, $\Omega _{\text{MW}}=2\pi \times  7.0\,\text{MHz} $, and the amplitude modulation depth of $\Omega _{\text{MW}}$ is 5\%-100\%. The inset shows that the EIT spectrum with $\Omega_{\text{MW}}=0$ and the line width of the EIT is about 4 MHz.}
\end{figure}

We further investigate the effects of MD on the dispersion signal, as well as $\Delta f_{\text{zeros}}$ and $\Delta f_{\text{sho}}$. The dispersion signals with different MD are calculated and shown in Figure \ref{fig:5}, the $\Omega_{\text{MW}} = 2\pi \times 2.0$ MHz is set lower enough that the EIT-AT splitting can not be distinguished as shown in the inset of Figure \ref{fig:5}. The corresponding dispersion signals have two zero-crossing points and a splitting symmetric double-peak. As the MD increases from 10\% to 100\%, the splitting symmetric double-peak becomes higher and clearer. $\Delta f_{\text{sho}}$ as well as $\Delta f_{\text{zeros}}$ becomes smaller as the MD increases, due to the equivalent MW Rabi frequency becomes smaller. The relationships of $\Delta f_{\text{zeros}}/\Omega_{\text{MW}}$, $\Delta f_{\text{sho}}/\Omega_{\text{MW}}$ with MD are shown in Figure \ref{fig:6}. When the MD of $\Omega_{\text{MW}}$ is below 60\%, $\Delta f_{\text{zeros}}/\Omega_{\text{MW}}$, $\Delta f_{\text{sho}}/\Omega_{\text{MW}}$ show a good linear relationship with the MD, and as the MD becomes larger, $\Delta f_{\text{zeros}}/\Omega_{\text{MW}}$ and $\Delta f_{\text{sho}}/\Omega_{\text{MW}}$ become smaller. This is due to the fact that the MD of $\Omega_{\text{MW}}$ becomes larger, leading to a decrease in the effective MW Rabi frequency, and $\Delta f_{\text{zeros}}/\Omega_{\text{MW}}$ and $\Delta f_{\text{sho}}/\Omega_{\text{MW}}$ also indirectly become smaller. When the MD of $\Omega_{\text{MW}}$ is greater than 60\%, the MW effective Rabi frequency constantly approaches the EIT linewidth without MW (The EIT spectral line width in the lower left corner of Figure~\ref{fig:6} is about 4 MHz), then $\Delta f_{\text{zeros}}/\Omega_{\text{MW}}$ and $\Delta f_{\text{sho}}/\Omega_{\text{MW}}$ are nonlinearly related to the MD. The results confirm that $\Delta f_{\text{zeros}}$ and $\Delta f_{\text{sho}}$ can be clearly obtained over a large range of MD, and $\Delta f_{\text{zeros}}/\Omega_{\text{MW}}$, $\Delta f_{\text{sho}}/\Omega_{\text{MW}}$ are constant when the modulation depth is fixed.

\begin{figure}
\includegraphics[scale=0.8]{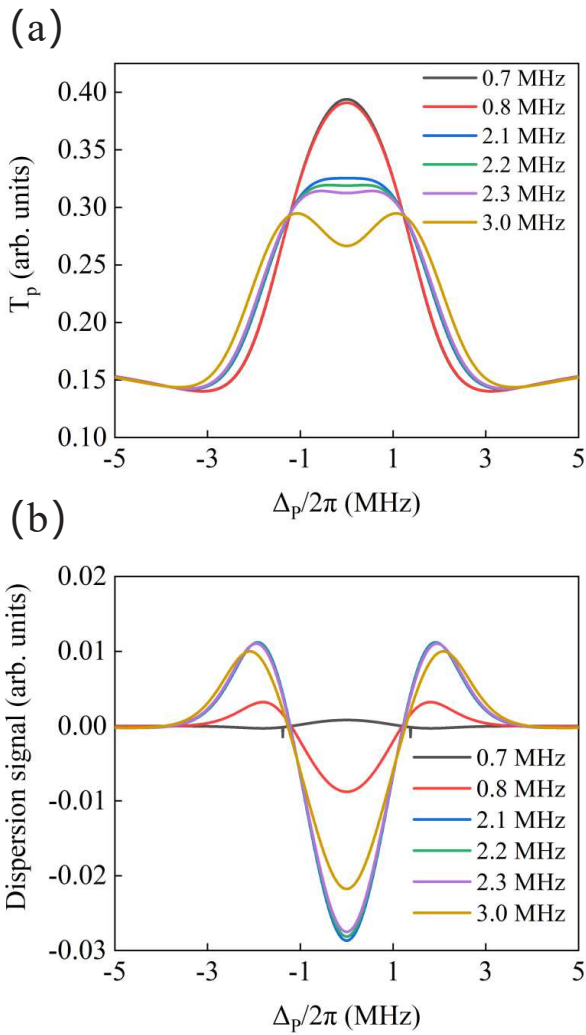}
\caption{\label{fig:3} Numerical simulation results of EIT-AT splitting spectra at MW field Rabi frequencies $\Omega_{\text{MW}}$ between 0.7 MHz and 3 MHz.  (a) EIT-AT spectra without modulation;  (b) Corresponding dispersion error signal with modulation depth MD=100\%.}
\end{figure}

We calculate the dispersion signal of EIT-AT at different $\Omega_{\text{MW}}$ to discuss the feasibility of $\Delta f_{\text{sho}}$ to measure weak fields. Figure~\ref{fig:3} shows the numerical simulation results of EIT-AT splitting with MW field Rabi frequencies $\Omega_{\text{MW}}$ from 0.7 MHz to 3 MHz. Where Figure~\ref{fig:3}(a) represents the EIT-AT splitting spectra at $\Omega _{\text{MW}}$ without amplitude modulation. $\Delta f_{\text{m}}$ becomes smaller when $\Omega _{\text{MW}}$ decreases, and $\Delta f_{\text{m}}$ becomes indistinguishable when $\Omega _{\text{MW}}$ is less than 2.2 MHz. Figure~\ref{fig:3}(b) represents the dispersion error signal after amplitude modulation of $\Omega _{\text{MW}}$. When $\Omega _{\text{MW}}$ is much smaller than 2.2 MHz, $\Delta f_{\text{sho}}$ can still be observed. 

\begin{figure}
\includegraphics[scale=0.35]{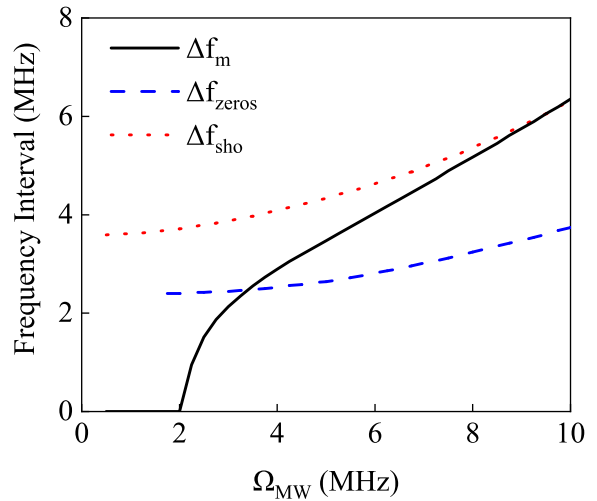}
\caption{\label{fig:4} Calculation results with $\Delta f_{\text{m}}$, $\Delta f_{\text{zeros}}$ and $\Delta f_{\text{sho}}$ with respect to $\Omega _{\text{MW}}$. The solid black line represents the original EIT-AT splitting interval $\Delta f_{\text{m}}$; the blue dashed line represents the interval of zero crossing of dispersion signals $\Delta f_{\text{zeros}}$; the red dotted line represents the dispersion signal shoulder interval $\Delta f_{\text{sho}}$. The calculation parameters are $\Omega_{\text{MW}}=2\pi \times  0.7\,\text{MHz}\sim2\pi \times  10\,\text{MHz}$, $\Omega_{\text{p}}=2\pi \times  1.0\,\text{MHz} $, $\Omega_{\text{c}}=2\pi \times  4.0\,\text{MHz} $ }
\end{figure}

The relationship between $\Delta f_{\text{m}}$, $ \Delta f_{\text{zeros}}$ and $\Delta f_{\text{sho}}$ for $\Omega _{\text{MW}}$ between 0.7-10 MHz is shown in Figure \ref{fig:4}. As shown in the black solid line in Figure \ref{fig:4}, $\Delta f_{\text{m}}$ decreases linearly with $\Omega _{\text{MW}}$ decreasing, but when the $\Omega _{\text{MW}}$ is less than 3.4 MHz, the linear characteristic of $\Delta f_{\text{m}}$ disappears and finally becomes indistinguishable with $\Omega _{\text{MW}}$ = 2.0 MHz.
The blue dashed line in Figure \ref{fig:4}, $\Delta f_{\text{zeros}}$ decreases monotonically with $\Omega _{\text{MW}}$ decreasing, and $\Delta f_{\text{zeros}}$ becomes unchanging when $\Omega _{\text{MW}}$ $\textless$ 1.5 MHz. The red dotted line in Figure \ref{fig:4}, $\Delta f_{\text{sho}}$ decreases monotonically with $\Omega _{\text{MW}}$ decreasing, and $\Delta f_{\text{sho}}$ becomes unchanging $\Omega _{\text{MW}}$ when $\Omega _{\text{MW}}$ is 0.8 MHz. The results show that $\Delta f_{\text{sho}}$ has a significant resolution advantage over $\Delta f_{\text{m}}$ and $\Delta f_{\text{zeros}}$ when measures weak MW weak fields.

\section{EXPERIMENTAl SYSTEM AND METHODS}
\begin{figure}
\includegraphics[scale=0.33]{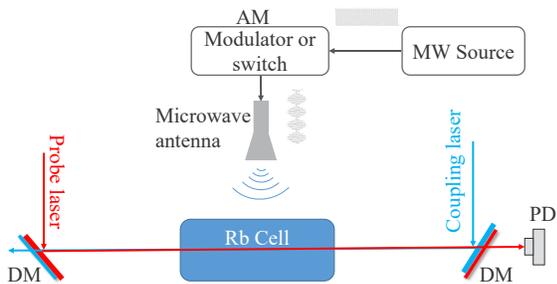}
\caption{\label{fig:7} The experimental setup for measuring the MW strength by amplitude modulation method. The red and blue lines with arrows represent the probe laser and coupling laser, respectively. A modulator or switch is used to modulate MW amplitude. DM is the dichroic mirror, and PD is the photodetector.}
\end{figure}

\begin{figure}
\includegraphics[scale=0.4]{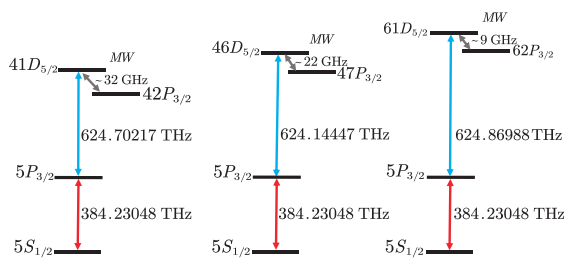}
\caption{\label{fig:8} Selection of energy levels of rubidium for measuring MW with the frequency of 31.6 GHz, 22.1 GHz, and 9.2 GHz, respectively. In order to clearly show the energy levels, the energy level positions are not drawn strictly at real intervals.}
\end{figure}

We have experimentally verified the above proposal in a $^{87}Rb$ vapor cell at room temperature. The schematic diagram of the experimental system is shown in Figure \ref{fig:7}. The diameter and length of the Rb vapor cell at room temperature were 25 mm and 75 mm respectively. In the four energy levels proposed in Figure~\ref{fig:1}, when the MW field frequency to be measured is 31.6 GHz by $^{87}Rb$ atoms, the four energy levels $\ket{1}$, $\ket{2}$, $\ket{3}$, and $\ket{4}$ in the theoretical model can be selected as $5S_{1/2}$, $5P_{3/2}$, $41D_{5/2}$, and $42P_{3/2}$, respectively, as shown in Figure~\ref{fig:8}. Where the probe laser ($\lambda_{\text{p}}$ $\sim$ 780 nm, DL100, Toptica) is generated by a tunable diode laser whose frequency is locked at the transition of $^{87}Rb$ $5S_{1/2}\rightarrow 5P_{3/2}$. The coupling light ($\lambda_{\text{c}}$ $\sim$ 480nm, TA-SHG Pro, Toptica) is generated by a frequency doubling diode laser, its frequency is locked at the resonant transition of $^{87}Rb$ $5P_{3/2}\rightarrow 41D_{5/2}$ by the Zeeman modulation EIT method \cite{Jia_20}. The probe and coupling lasers overlap and transmit in reverse on the central axis of the Rb cell. The MW amplitude modulation is realized by inputting TTL or sin signal into the amplitude modulation port of the microwave source, and finally the modulated MW is irradiated on the vapor cell through a microwave antenna (LB-28-15-C-KF). A photodetector (PDA36A2, Thorlabs) is used to receive the transmission signal of the probe laser through the $^{87}Rb$ vapor cell and then record by an oscilloscope.

\begin{figure*}
\includegraphics[scale=0.7]{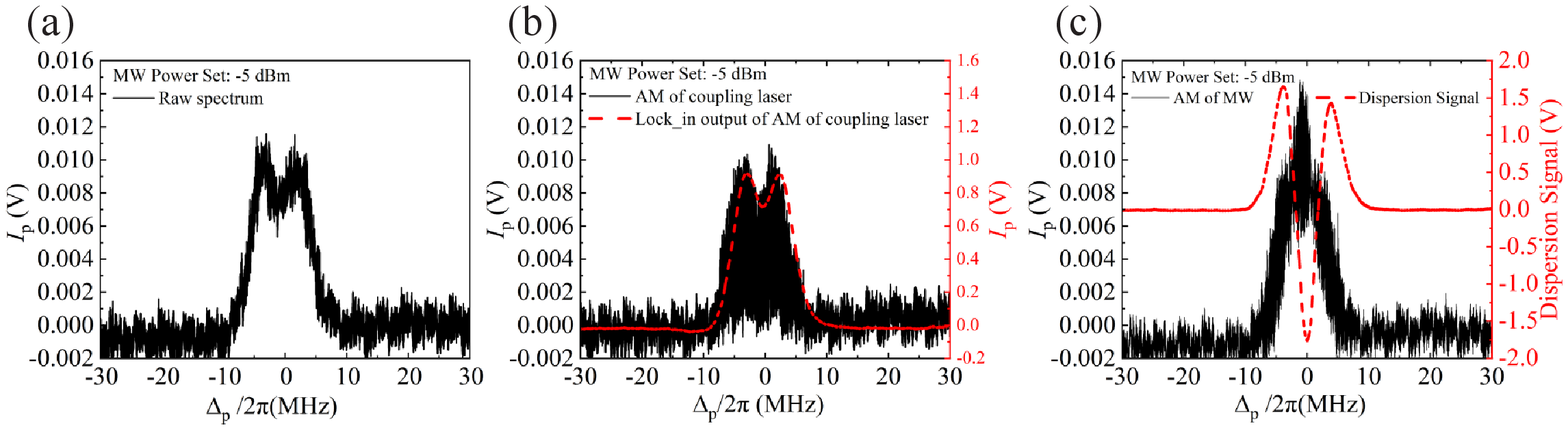}
\caption{\label{fig:9} Experimental examples of the spectra: (a) original EIT-AT splitting spectrum without AM modulation, (b) the coupling laser amplitude modulation spectrum, and (c) the MW amplitude modulation spectrum, respectively. The amplitude modulation frequency is 10 kHz and the amplitude modulation depth is 100\%.}
\end{figure*}

The MW field in the experiment was provided by an MW source (E8257D, Keysight Technologies) and radiated by a horn antenna of the corresponding frequency (see below for specific models) to the Rb vapor cell. MW fields at three different frequencies (9.2 GHz, 22.1 GHz, and 31.6 GHz, respectively) are measured in this work. The schematic diagram of the energy level is shown in Figure \ref{fig:8}. For the MW field at 31.6 GHz, the coupling laser coupling energy level $5P_{3/2}$ $\rightarrow$ $41D_{5/2}$ and MW field coupling energy level $41D_{5/2}$ $\rightarrow$ $42P_{3/2}$, corresponding to a resonant transition dipole moment of $3309.549\,a_0e$; For MW field with 22.1 GHz, the coupling laser coupling energy level $5P_{3/2}$ $\rightarrow$ $46D_{5/2}$ and MW field coupling energy level $46D_{5/2}$ $\rightarrow$ $47P_{3/2}$, corresponding to a resonant transition dipole moment of $4195.706\,a_0e$; For the MW field with 9.2 GHz, the coupling laser coupling energy level $5P_{3/2}$ $\rightarrow$ $61D_{5/2}$ and MW field coupling energy level $61D_{5/2}$ $\rightarrow$ $62P_{3/2}$, corresponding to a resonant transition dipole moment of $7483.291\,a_0e$, where $a_0$ is the Bohr radius and $e$ is the elementary charge; The MW antenna is placed as far away as possible from the Rb vapor to match the far-field conditions. The MW field, probe laser, and coupling laser are all linearly polarized. A sin signal is used to modulate the MW source. The range of modulation frequency in this work can be set from 1 to 99 kHz, and the MD of the MW can be changed from 0\% to 100\%. The Rydberg EIT-AT spectra are obtained by scanning the probe laser frequency near the EIT resonance using an Acousto-optic modulator (AOM). The modulated EIT signal is sent to a lock-in amplifier (LI5640, NF Corporation) for demodulation, and then the dispersion signal can be obtained.

\section{RESULTS AND DISCUSSION}
We first studied the Rydberg EIT-AT spectrum obtained without modulation of the MW and the dispersion signal obtained by modulation. The experimental results are shown in Figure \ref{fig:9}. Figure 9(a) shows the original EIT-AT splitting spectrum where the signal-to-noise ratio(SNR) is relatively poor. The solid black line in Figure 9(b) indicates the EIT-AT spectrum obtained by modulating the coupling laser amplitude, the red dashed line indicates the signal demodulated by the lock-in amplifier, and the SNR of the demodulated spectrum is improved by 17.5 dB compared with the original EIT-AT spectrum in Figure 9(a). The solid black line in Figure 9(c) represents the EIT-AT spectrum obtained by modulating the MW amplitude, and the red dashed line represents the dispersion signal obtained by MW amplitude modulation. The SNR of the dispersion signal is improved by 20.2 dB compared with the original EIT-AT spectrum in Figure 9(a). It can be seen that the dispersion signal has two zero-crossing points whose interval is defined as $\Delta f_{\text{zeros}}$. Xiu-bin Liu \emph{et al}. have demonstrated that $\Delta f_{\text{zeros}}$ extends the lower limit of MW measurement by a factor of 6 compared to the conventional 
EIT-AT splitting method \cite{liu_2021}. More interesting, there are two positive maxima of the dispersive signal whose interval is defined as $\Delta f_{\text{sho}}$, $\Delta f_{\text{sho}}$ is larger than $\Delta f_{\text{zeros}}$ which could reflect the weaker $E_{\text{MW}}$. Compared with the former and based on the theoretical model in Sec. II, the lower limit of the MW electric field that can be measured by this method is also lower.

We then performed a detailed analysis of the spectra obtained by the conventional method \cite{4} and the AM modulation method at 31.6 GHz. For the conventional method, when the MW electric field is continuously varied between 1.22 and 15.35 mV/cm, a series of spectral signals corresponding to the EIT-AT splitting are obtained as shown in Figure 10(a). The experimental results show that $\Delta f_{\text{m}}$ becomes indistinguishable as $E_{\text{MW}}$ gradually decreases to 4.85 mV/cm. Even for $E_{\text{MW}}$=4.85 mV/cm, we cannot read $\Delta f_{\text{m}}$ directly from the graph, but can only get it indirectly by spectrum analysis. The reason is that $E_{\text{MW}}$ is too small, the two dressed-level intervals by MW are smaller than the EIT linewidth $\Gamma _{\text{EIT}}$ which results that the Rydberg EIT-AT splitting cannot be directly observed. Furthermore, when $E_{\text{MW}}$ is smaller than 1.22 mV/cm, $\Delta f_{\text{m}}$ cannot be obtained even by spectrum analysis. In contrast, as shown in Figure 10(b), $\Delta f_{\text{sho}}$ is still very distinguishable in dispersion signal with $E_{\text{MW}}$ is smaller than 1.22 mV/cm. We find that $\Delta f_{\text{sho}}$ are still present under the weak field where the EIT-AT splitting is indistinguishable, which confirms that $\Delta f_{\text{sho}}$ can be used to measure and characterize the weak MW electric field strength. More interesting, $\Delta f_{\text{sho}}$ is still clearly distinguished when the interval $ \Delta f_{\text{zeros}}$ does not exist with the much weaker MW field as shown in Figure 10(b).

\begin{figure}
\includegraphics[scale=0.5]{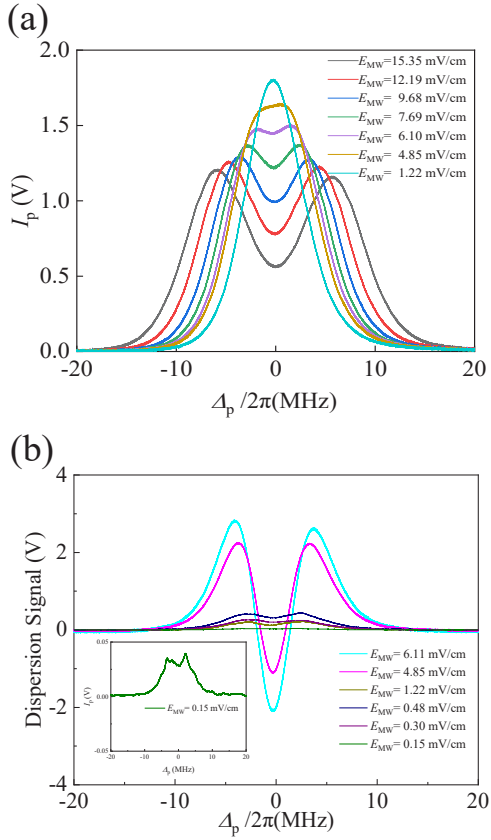}
\caption{\label{fig:10} Experimental conventional Rydberg EIT-AT splitting spectra and the dispersion signal obtained by AM modulation of MW. (a) EIT-AT splitting spectra with the $E_{\text{MW}}$ reduced from 15.35 mV/cm to 1.22 mV/cm. (b) Dispersion signal with the $E_{\text{MW}}$ reduced from 6.11 mV/cm to 0.15 mV/cm. The inset shows that the dispersion signal with $E_{\text{MW}}$ = 0.15 mV/cm. In the experiment, the modulation frequency of MW is 10 kHz and the modulation depth is 100\%. 
}
\end{figure}

\begin{figure}
\includegraphics[scale=0.33]{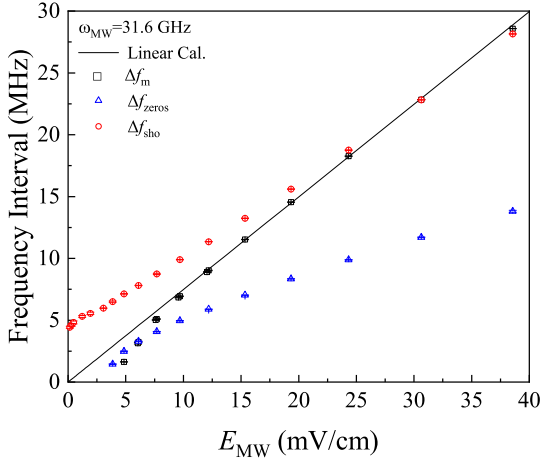}
\caption{\label{fig:11} Experimental results of $ \Delta f_{\text{m}}$, $ \Delta f_{\text{zeros}}$ and $ \Delta f_{\text{sho}}$ versus $ E_{\text{MW}}$ at MW frequency of 31.6 GHz. The black hollow squares represent results of the traditional EIT-AT splitting method by $ \Delta f_{\text{m}}$. The blue hollow triangles represents the experimental results of the interval of zero crossing of dispersion signals method by $ \Delta f_{\text{zeros}}$. The red hollow circles represent the experimental results of the dispersion signal shoulder interval method by $ \Delta f_{\text{sho}}$. The modulation frequency of MW is 10 kHz and the modulation depth is 100\%.}
\end{figure}

Next, we discuss in detail the experimental results of the $\Delta f_{\text{m}}$ method, $\Delta f_{\text{zeros}}$ method, and $ \Delta f_{\text{sho}}$ method at an MW frequency of 31.6 GHz. The $\Delta f_{\text{m}}$, $\Delta f_{\text{zeros}}$ and $ \Delta f_{\text{sho}}$ corresponding to different $ E_{\text{MW}}$ can be further obtained from Figure \ref{fig:10}, and the results are shown in Figure \ref{fig:11}. The black hollow square dots indicate the variation of the EIT-AT splitting interval $ \Delta f_{\text{m}}$ with the MW field $ E_{\text{MW}}$ by the conventional EIT-AT splitting method, which exhibits a linear feature when $ E_{\text{MW}}$ $\textgreater$ 7.69 mV/cm. As $ E_{\text{MW}}$ decreases, the $\Delta f_{\text{m}}$ becomes smaller than the linear prediction by Eq.~(\ref{eq1}) and can not be observed when $ E_{\text{MW}}$ $\textless$ 4.854 mV/cm. For $ \Delta f_{\text{zeros}}$ mothed as shown as the blue hollow triangles in Figure~\ref{fig:11}, $ \Delta f_{\text{zeros}}$  can not be observed with $ E_{\text{MW}}$  <3.856 mV/cm in this work. Note that the improvement by $ \Delta f_{\text{zeros}}$ is not as well as the results of Xiu-bin Liu \emph{et al} \cite{liu_2021}, this is because the experimental parameters in our experiment were optimized to obtain clearer $ \Delta f_{\text{sho}}$ resolution instead of $ \Delta f_{\text{zeros}}$. For $ \Delta f_{\text{sho}}$ mothed as shown as the red hollow circles in Figure~\ref{fig:11}, $ \Delta f_{\text{sho}}$ is still clearly observed even when $ E_{\text{MW}}$ = 0.152 mV/cm, and the monotonic relationship between $ E_{\text{MW}}$ and $ \Delta f_{\text{sho}}$ are always kept. As the results with MW frequency of 31.6 GHz, $ \Delta f_{\text{sho}}$ in this work is better than $ \Delta f_{\text{zeros}}$ method \cite{liu_2021}, and the lower limit of $ E_{\text{MW}}$ is even 32 times improved than that of the conventional EIT-AT splitting method \cite{4}. 

\begin{figure}
\includegraphics[scale=0.33]{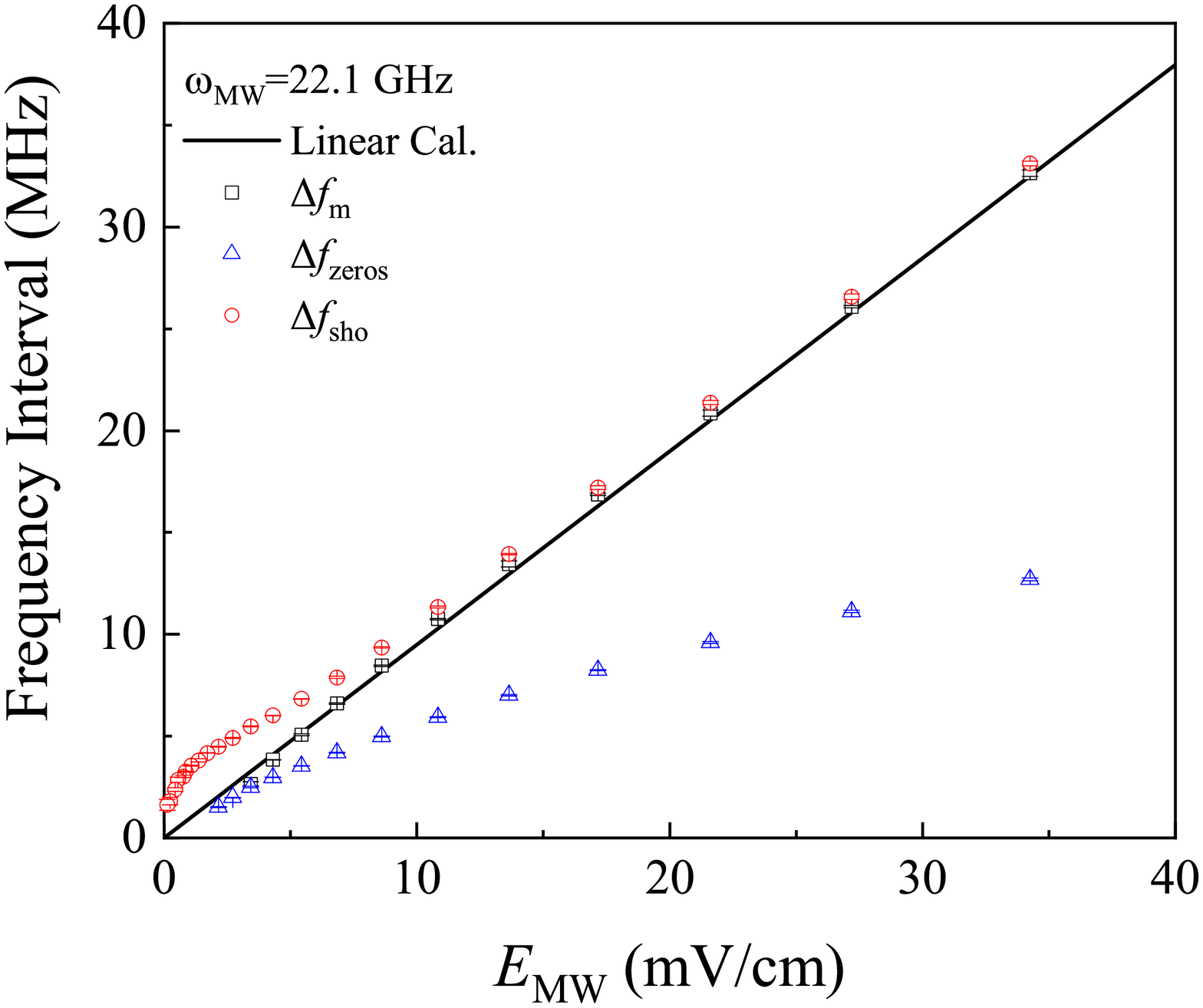}
c\caption{\label{fig:12} Experimental results of $ \Delta f_{\text{m}}$, $ \Delta f_{\text{zeros}}$ and $ \Delta f_{\text{sho}}$ versus $ E_{\text{MW}}$ at MW frequency of 22.1 GHz. The black hollow squares represent results of the traditional EIT-AT splitting method by $ \Delta f_{\text{m}}$. The blue hollow triangles represents the experimental results of the interval of zero crossing of dispersion signals method by $ \Delta f_{\text{zeros}}$. The red hollow circles represent the experimental results of the dispersion signal shoulder interval method by $ \Delta f_{\text{sho}}$. The modulation frequency of MW is 10 kHz and the modulation depth is 100\%.}
\end{figure}

We discuss in detail the experimental results of the $\Delta f_{\text{m}}$ method, $\Delta f_{\text{zeros}}$ method, and $ \Delta f_{\text{sho}}$ method at an MW frequency of 22.1 GHz, the results are shown in Figure \ref{fig:12}. The black hollow square dots indicate the variation of the EIT-AT splitting interval $ \Delta f_{\text{m}}$ with the MW field $ E_{\text{MW}}$ by the conventional EIT-AT splitting method, which exhibits a linear feature when $ E_{\text{MW}}$ $\textgreater$ 4.311 mV/cm. As $ E_{\text{MW}}$ decreases, the $\Delta f_{\text{m}}$ becomes smaller than the linear prediction by Eq.~(\ref{eq1}) and can not be observed when $ E_{\text{MW}}$ $\textless$ 3.425 mV/cm. For $ \Delta f_{\text{zeros}}$ mothed as shown as the blue hollow triangles in Figure~\ref{fig:12}, $ \Delta f_{\text{zeros}}$  can not be observed with $ E_{\text{MW}}$  <2.161 mV/cm in this work. For $ \Delta f_{\text{sho}}$ mothed as shown as the red hollow circles in Figure~\ref{fig:12}, $ \Delta f_{\text{sho}}$ is still clearly observed even when $ E_{\text{MW}}$ = 0.136 mV/cm, and the monotonic relationship between $ E_{\text{MW}}$ and $ \Delta f_{\text{sho}}$ are always kept. As the results with MW frequency of 22.1 GHz, $ \Delta f_{\text{sho}}$ in this work is better than $ \Delta f_{\text{zeros}}$ method \cite{liu_2021}, and the lower limit of $ E_{\text{MW}}$ is even 24 times improved than that of the conventional EIT-AT splitting method \cite{4}.

\begin{figure}
\includegraphics[scale=0.33]{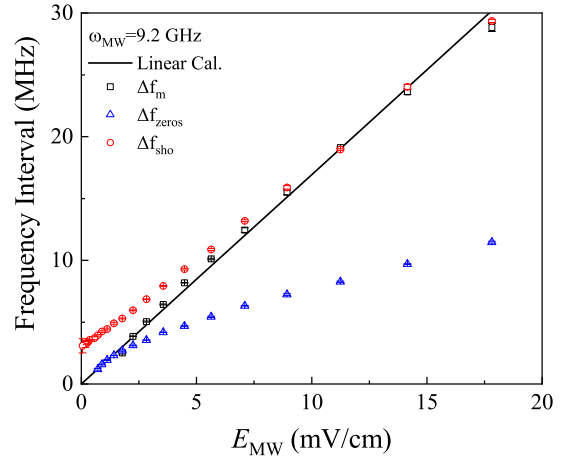}
\caption{\label{fig:13} Experimental results of $ \Delta f_{\text{m}}$, $ \Delta f_{\text{zeros}}$ and $ \Delta f_{\text{sho}}$ versus $ E_{\text{MW}}$ at MW frequency of 9.2 GHz. The black hollow squares represent the results of the traditional EIT-AT splitting method by $ \Delta f_{\text{m}}$. The blue hollow triangles represent the experimental results of the interval of zero crossing of dispersion signals method by $ \Delta f_{\text{zeros}}$. The red hollow circles represent the experimental results of the dispersion signal shoulder interval method by $ \Delta f_{\text{sho}}$. The modulation frequency of MW is 10 kHz and the modulation depth is 100\%.}
\end{figure}

We further discuss and analyze the measurement results of the three methods at 9.2 GHz MW frequency in order to verify the effectiveness of the proposed method for MW measurement at 1 to 40 GHz. The results are shown in Figure \ref{fig:13}. The process of analysis is similar to the above, the results with MW frequency of 9.2 GHz show that the lower limit of $ E_{\text{MW}}$ = 0.056 mV/cm is even 31 times improved than that of the conventional EIT-AT splitting method \cite{4}. 

Finally, the comparison of the minimum MW strength $E_{\text{MW}}$ measured by different methods is shown in Table~\ref{tab:table1}, with MW frequencies of 31.6 GHz, 22.1 GHz, and 9.2 GHz, respectively. $E_{\text{MW}}{(}\Delta f_{\text{m}}{)}_{\text{min}}$, $E_{\text{MW}}{(}\Delta f_{\text{zeros}}{)}_{\text{min}}$ and $E_{\text{MW}}{(}\Delta f_{\text{sho}}{)}_{\text{min}}$ are the lower limits for measuring the distinguishable MW field strength at specific MW frequencies. When the MW frequency is 9.2 GHz, the lower limit of the $E_{\text{MW}}$ can be measured is 0.056 mV/cm, which is much smaller than that when the MW frequency is 22.1 GHz and 31.6 GHz. The quantum number of the two adjacent Rydberg levels is relatively high when the MW frequency is 9.2 GHz, which makes the electric dipole moment between the two Rydberg levels large. According to Eq. (\ref{eq4}), for the same spectral splitting interval, the greater the electric dipole moment, the lower limit of the measurable microwave field.  Since then, this work has verified that $\Delta f_{\text{sho}}$ method has advantages over traditional methods and $\Delta f_{\text{zeros}}$ method in MW field measurement under the experimental conditions of MW frequencies of 31.6 GHz, 22.1 GHz, and 9.2 GHz. This is also predicted in the theoretical part of Sec. II.

\begin{table*}
\caption{\label{tab:table1} Comparison of the minimum MW strength $E_{\text{MW}}$  
measured by different methods, with MW frequencies of 31.6 GHz, 22.1 GHz, and 9.2 GHz, respectively.}
\begin{ruledtabular}
\begin{tabular}{ccccc}
 Frequency&$E_{\text{MW}}{(}\Delta f_{\text{m}}{)}_{\text{min}}$&$E_{\text{MW}}{(}\Delta f_{\text{zeros}}{)}_{\text{min}}$&$E_{\text{MW}}{(}\Delta f_{\text{sho}}{)}_{\text{min}}$
\\ \hline
 31.6\,GHz&$4.854\pm 0.024\,\text{mV/cm}$&$3.856\pm 0.032\,\text{mV/cm}$ &$0.152\pm 0.002\,\text{mV/cm}$ \\
 22.1\,GHz&$3.425\pm 0.064\,\text{mV/cm}$&$2.161\pm 0.012\,\text{mV/cm}$&$0.136\pm0.011\,\text{mV/cm}$\\
 9.2\,GHz&$1.782\pm 0.033\,\text{mV/cm}$&$0.709\pm 0.018\,\text{mV/cm}$&$0.056\pm 0.005\,\text{mV/cm}$\\
\end{tabular}
\end{ruledtabular}
\end{table*}

\section{CONCLUSION}
We have theoretically and experimentally studied the dispersion signal of the Rydberg atomic electromagnetic induction transparent Autler-Townes splitting spectrum obtained using amplitude modulation of the MW field in detail. Theoretically, we first demonstrate the process of microwave electric field amplitude modulation and demodulation to obtain the dispersion signals by numerical simulation. Compared with the traditional EIT-AT splitting interval $\Delta f_{\text{m}}$, the dispersion signal provides two spectral intervals, namely the interval of the two zero-crossing points $\Delta f_{\text{zeros}}$  and the interval of two positive maxima $\Delta f_{\text{sho}}$. Then we study the effect of modulation depth on the dispersion signals in detail, and the results show that the greater the modulation depth, the more pronounced the characteristics of zero-crossing points and maxima in the dispersion signal. When the modulation depth is fixed, the ratio of $\Delta f_{\text{zeros}}/\Omega_{\text{MW}}$ or $\Delta f_{\text{sho}}/\Omega_{\text{MW}}$ is fixed, which proves that $\Delta f_{\text{zeros}}$ and $\Delta f_{\text{sho}}$ can be used to characterize microwave electric field strength $E_{\text{MW}}$. The traditional EIT-AT spectra and the corresponding dispersion signals at 100\% modulation depth at different $\Omega_{\text{MW}}$ are theoretically calculated. By comparing $\Delta f_{\text{m}}$, $\Delta f_{\text{zeros}}$, and $\Delta f_{\text{sho}}$ with $\Omega_{\text{MW}}$ in detail, the results show that $\Delta f_{\text{sho}}$ can characterize much weaker microwave electric fields than $\Delta f_{\text{m}}$, $\Delta f_{\text{zeros}}$. Experimentally, we have studied the relationship of MW field strength $E_{\text{MW}}$ and $\Delta f_{sho}$ in a rubidium vapour cell at the MW frequencies of 31.6 GHz, 22.1 GHz, and 9.2 GHz, respectively. The results show that the microwave electric field amplitude modulation method improves the spectral signal-to-noise ratio, and more interesting $\Delta f_{\text{sho}}$ can be used to characterize the much weaker $E_{\text{MW}}$ than $\Delta f_{\text{m}}$ and $\Delta f_{\text{zeros}}$, the minimum $E_{\text{MW}}$ measured by $\Delta f_{\text{sho}}$ is about 30 times smaller than the conventional EIT-AT splitting method. As an example, the minimum $E_{\text{MW}}$ at 9.2 GHz that can be characterized by $\Delta {\text{sho}}$ is 0.056 mV/cm, which is the minimum value we know that can be characterized by frequency interval using vapour cell without adding any auxiliary fields. The proposed method can further reduce the weak limit of $E_{\text{MW}}$ measured by spectral frequency interval, which is important in the direct measurement of weak $E_{\text{MW}}$.

\nocite{*}
\bibliography{apssamp}

\end{document}